\documentclass{article}

\usepackage[english]{babel}

\usepackage[a4paper,top=2cm,bottom=2cm,left=3cm,right=3cm,marginparwidth=1.75cm]{geometry}

\usepackage{amsmath}
\usepackage{graphicx}
\usepackage[backend=biber,style=chem-rsc,sorting=none]{biblatex}
\usepackage{mhchem}
\usepackage{booktabs} 
\usepackage{float}
\usepackage{siunitx} 
\usepackage{pdfpages}

\title{Prediction of Novel Li–Ag(II)–F Compounds using Evolutionary Algorithms}
\author{Katarzyna Kuder$^{a,*}$, Wojciech Grochala$^{a,**}$}

\date{}

\begin{document}
\maketitle
$^{a}$ Centre of New Technologies, University of Warsaw, S. Banacha 2c, 02-097 Warsaw, Poland

\begin{center}
$^{*}$ k.kuder@cent.uw.edu.pl

$^{**}$ w.grochala@cent.uw.edu.pl
\end{center}

\begin{center}
\textit{This article is dedicated to Piotr J.  Leszczyński at his 60$^{th}$ birthday.}
\end{center}

\begin{abstract}

This work provides a theoretical exploration of the thermodynamic stability and magnetic behaviour of previously unknown ternary Li--Ag(II)--F compounds. Convex-hull analysis shows that all predicted structures lie slightly above the LiF + AgF$_2$ decomposition line, indicating a natural tendency toward phase separation; nevertheless, their negative formation energies relative to AgF, LiF, and F$_2$/F suggest that alternative synthetic pathways may be feasible for these compounds.
All studied structures show preference for antiferromagnetic ground state. Notably, the triclinic LiAgF$_3\_2$ is predicted to exhibit an exceptionally large superexchange constant ($J=~-358$~meV) within $[{\rm Ag_2F_7}]$ dimers, placing it above the strongest known magnetic exchange interactions reported to date.

\end{abstract}

\section{Introduction}

The properties of silver(II) fluoride has been thoroughly documented since the 1960s through the 1970s \cite{1,2}. This compound is distinguished by the presence of covalent Ag-F bonds \cite{3} within layered structural motifs and pronounced antiferromagnetic interactions  \cite{4,5,6,8}. Owing to its high reactivity, primarily due to facile reduction to silver(I), silver(II) fluoride has been considered more of a curiosity than a subject warranting thorough investigation. This perception began to change in the 2000s when the interest in examining Ag(II)-containing systems was rekindled. In this period, such systems were noticed to be analogues to cuprates and were thoroughly investigated \cite{7,9}. One of the most striking outcomes of these studies was the realization that Ag(II) fluorides may exhibit record-breaking magnetic superexchange constants, which are comparable (and even greater) than those in copper(II) oxides \cite{20,21,62}.

A  variety of silver(II)-fluoride compounds are known, many of which are binary and exhibit mixed-valence character, such as \ce{Ag2F3}, \ce{Ag3F4}, \ce{Ag2F5}, and \ce{Ag3F8} \cite{22,9,10,11,12,13}. In addition, numerous ternary or more complex compounds containing silver, fluorine, and other elements have been reported, further expanding the structural and chemical diversity of Ag(II)--fluoride systems \cite{9, 42, 43, 44}. Despite the extensive exploration of silver(II) fluorides, there is a notable absence of studies focusing on lithium–silver(II)–fluoride systems, even though related systems containing homologous sodium, potassium, caesium, and rubidium have been investigated both experimentally and theoretically \cite{15, 16, 46, 47, 48}. This  is particularly intriguing given the established interest in lithium fluoride (LiF) as a component in various applications, including optics and battery technologies \cite{17,18,19}. The combination of lithium fluoride with silver(II) fluoride could potentially lead to novel materials with unique properties, yet this area remains largely unexplored. To fill this gap, we initiated a systematic theoretical investigation of ternary systems containing lithium, silver(II) and fluorine using density functional theory (DFT). We performed a systematic exploration of ternary Li–Ag(II)–F systems using the XtalOpt evolutionary algorithm \cite{23, 24}, aimed at identifying energetically favourable structures across a broad range of stoichiometries.

This study provides the first theoretical insights into Li–Ag(II)–F systems, revealing their structural diversity, thermodynamic stability,  magnetic behaviour and their potential preparation pathways.

\section{Methods}

The structural screening was conducted using self-learning algorithms incorporated within the XtalOpt r.13.2 software \cite{23,24}. Structures were randomly generated and augmented by those derived from previous studies on systems containing related alkali metals $M^{(I)}Ag^{(II)}F_3$, at both ambient and elevated pressure conditions \cite{16}. The latter were altered by $M^+ \longrightarrow Li^+$ substitutions. Considering that lithium is considerably smaller than the other alkali metal cations, the high-pressure polymorphs were introduced into the initial pool of structures. Similarly to the methodology applied in \ce{Li_2AgF_4}, the study also involved the incorporation of modified structures into the initial pool \cite{49, 15}, with subsequent alterations executed accordingly. Unit cells containing 2, 4, or 6 formula units per cell were tested. For each stoichiometry and formula units per unit cell, usually 650 structures were generated, in which 100 of them  were initial starting structures.

The calculations were performed using VASP 5.4.4 (Vienna Ab initio Simulation Package) \cite{26,27,28}. The projector augmented-wave (PAW) method \cite{29,30}, with pseudopotentials generated within the PBE formalism, was employed to describe the interaction between valence electrons and ionic cores. The exchange–correlation potential was treated within the generalized gradient approximation (GGA) using the PBEsol functional \cite{31}.
The cut-off energy of the plane wave basis set was equal to 600 eV with a self-consistent-field convergence criterion of $10^{-7}$ eV.

 Numerous equilibrium structures were obtained. In the second stage, the most promising fifteen unique (lowest enthalpy)  structures obtained from preliminary screening were optimized in various ferromagnetic and antiferromagnetic models using the DFT+U for the proper description of localized d electrons in the Dudarev formulation \cite{32}, with on-site Coulomb and exchange parameters of U = 5.0 eV and J = 1.0 eV applied to the relevant d orbitals of Ag. The Hubbard $U$ correction $(DFT+U)$ is essential for an accurate description of the strongly correlated $4d$ electrons in Ag(II) \cite{5,20,21}.

Finally, the optimized structures were symmetrized using FINDSYM v.7.1.7 \cite{33,34}, and the superexchange constants were calculated.

\section{Results and Discussion}
\subsection{The crystal structures}

The structural search revealed a number of low-energy structures of LiAgF$_3$ and Li$_2$AgF$_4$. Table~\ref{T2} summarizes the calculated cell parameters for the lowest energy structures predicted together with their space groups. They are either monoclinic or triclinic and two belong to polar space groups (\textit{C}c and \textit{P}$2_1$). Importantly, all monoclinic polymorphs are unique structure types as none of them is analogous to any known polymorphs in the M$^{(I)}$Ag$^{(II)}$F$_3$ family (M=Na, K, Rb, Cs). The triclinic one, LiAgF$_3\_2$, however, is formally isotypic to AgCuF$_3$ and NaCuF$_3$ \cite{59}.

\begin{table}[h!]
\centering
\caption{Optimized unit cell parameters and crystallographic space groups for theoretically predicted Li--Ag--F structures of LiAgF$_3$ and Li$_2$AgF$_4$.}
\label{T2}
\begin{tabular}{lccccccc}
\hline
\textbf{Structure} & \textbf{Space group} & \textbf{$a$ (\AA)} & \textbf{$b$ (\AA)} & \textbf{$c$ (\AA)} & \textbf{$\alpha$ ($^{\circ}$)} & \textbf{$\beta$ ($^{\circ}$)} & \textbf{$\gamma$ ($^{\circ}$)} \\
\hline
LiAgF$_3\_1$ & $C$c & 3.410 & 14.628 & 4.999 & 90.0 & 93.8 & 90.0 \\
LiAgF$_3\_2$ & $P$$\bar{1}$ & 5.079 & 5.800 & 8.316 & 94.9 & 100.4 & 90.2 \\
Li$_2$AgF$_4\_1$         & $C$2/c & 20.986 & 5.739 & 5.009 & 90.0 & 99.4 & 90.0 \\
Li$_2$AgF$_4\_2$       & $P$$2_1/c$ & 5.481 & 5.799 & 5.032 & 90.0 & 107.1 & 90.0 \\
Li$_2$AgF$_4\_3$       & $P$$2_1$ & 3.294 & 5.205 & 9.778 & 90.0 & 94.6 & 90.0 \\
\hline
\end{tabular}
\end{table}

\begin{figure}[H]
    \centering
    \includegraphics[width=0.9\textwidth]{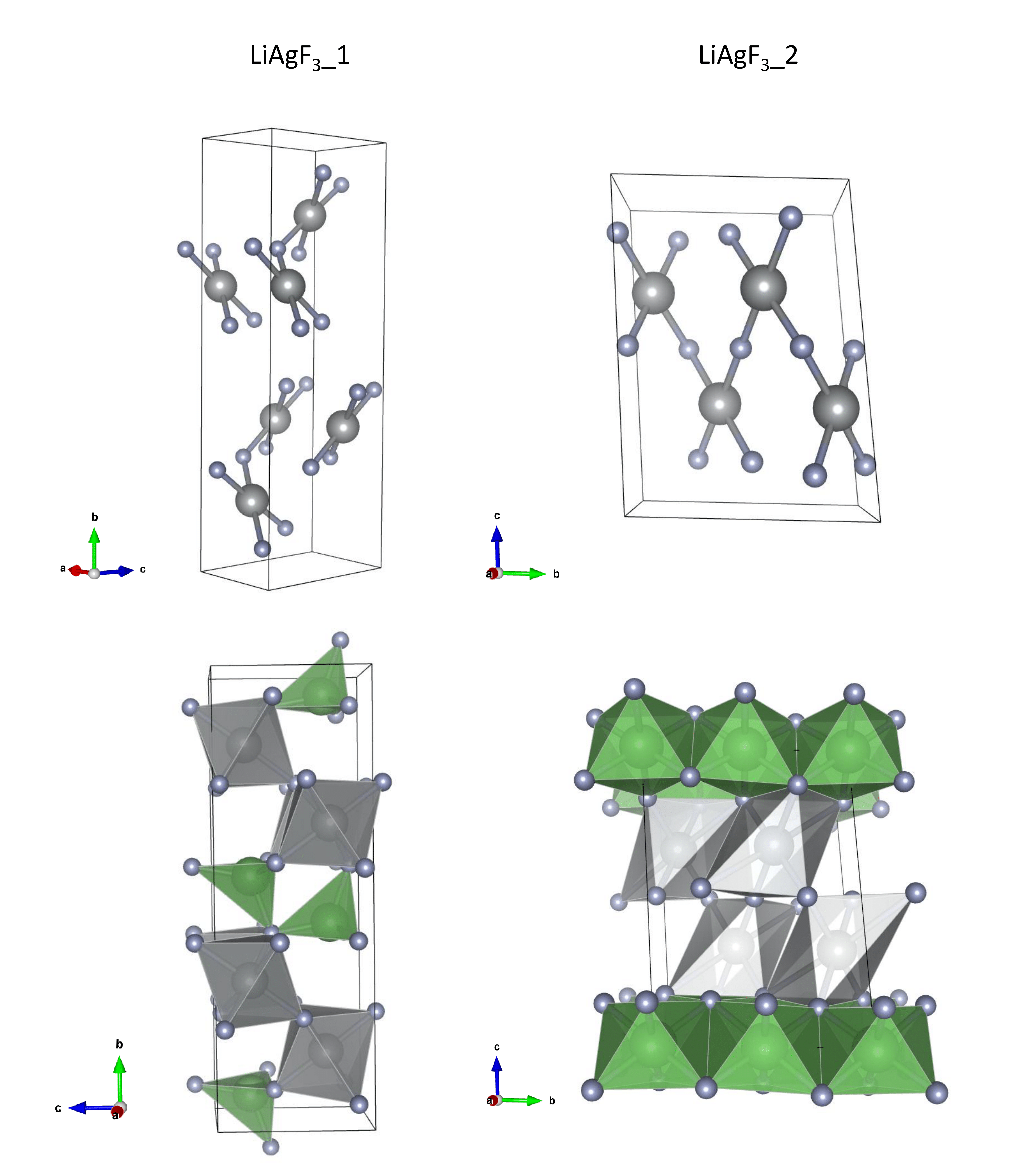}
    \caption{Crystal structures of the investigated LiAgF$_3$ structures obtained from DFT calculations. Large dark-gray spheres - Ag(II), small gray spheres - F, green spheres - Li. (top) Ag -- F sublattice, (middle) Li -- F sublattice, and (bottom) polyhedral coordination spheres are shown.}
    \label{F2}
\end{figure}

The lowest-energy LiAgF$_3\_1$ polymorph crystallises in the $C$c space group. The Ag(II) cation adopts an elongated octahedral environment, with four short Ag–F bonds (2.055\AA– 2.090\AA) and two longer ones around 2.50\AA. This geometry is very similar to that found for AgF$_2$ (four short bonds: 2.067\AA\, 2.071\AA\ and two long ones 2.588\AA) and it is  consistent with a Jahn–Teller distortion, a feature commonly observed in other Ag(II)–F systems \cite{36,37,38}. The Ag-F-Ag angle between the short bonds of adjacent Ag(II) cations is 136.9$^\circ$. The lithium cations exhibit disordered tetrahedral coordination, with Li-F bond lengths of 1.884\AA–1.932\AA, substantially shorter than those found in crystalline LiF (2.014\AA, with a regular octahedral environment \cite{39}). The $C$c polymorph has not yet been observed for other alkali metal -- Ag(II) fluorides. Its polar space group might indicate ferroelectric or even multiferroic properties, which are of interest for fluoride materials \cite{51, 52, 63}.

An alternative low-energy structure, LiAgF$_3\_2$, has approximately 3.7$\%$ smaller  volume than the ground state one and it belongs to the common triclinic centrosymmetric $P\overline{1}$ space group. Here, the Ag(II) environment is also elongated octahedral: the four shorter Ag–F bonds (2.012\AA–2.083\AA) are slightly shorter, whereas the two longer ones (2.821\AA–2.896\AA) are about 13.8$\%$ longer than in LiAgF$_3\_1$. Such behaviour originates from the plasticity of the Jahn-Teller effect-driven distortion of the first coordination sphere of Ag(II) \cite{53}. The corresponding Ag–F–Ag angles are 112.1$^\circ$ and 180$^\circ$ which leads to the formation of helix-like motifs in this structure. Lithium cation adopts an elongated octahedral coordination with two 1.985\AA, two 1.969\AA\ and two  2.124\AA\ bonds. An average Li -- F bond length is 2.026\AA\, thus not far from that for pristine LiF crystal (2.014\AA).

In contrast to other MAgF$_3$ compounds, which exhibit a distorted perovskite structure where the alkali metal atom substantially affects the degree of distortion \cite{20}, the LiAgF$_3$ structures explored in this work do not follow this trend. In both LiAgF$_3$ post-perovskite structures  (Figure \ref{F2}), kinked $[AgF_{2/1+2/2}]^{-}$ chains are observed. It is worth noting that for theoretically predicted high-pressure polymorphs of MAgF$_3$ \cite{14}, increasing pressure leads to a gradual tilting of the infinite $[AgF_6]^-$ chains. Thus, the LiAg$^{(II)}$F$_3$ polymorphs studied here turn out to be analogues of high-pressure phases of heavier alkali metal systems.

\begin{figure}[H]
    \centering
    \includegraphics[width=1\textwidth]{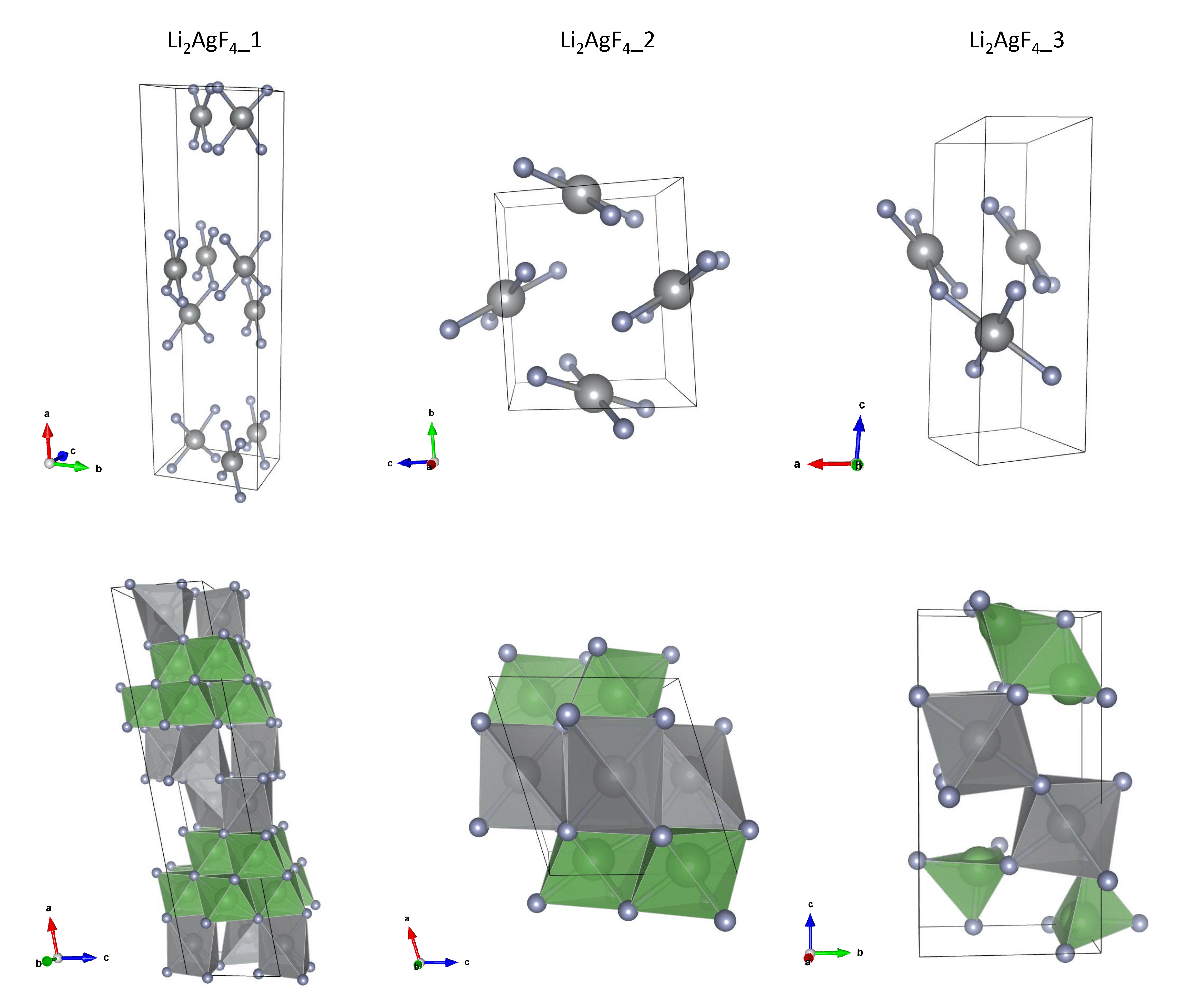} 
    \caption{Crystal structures of the investigated Li$_2$AgF$_4$ structures obtained from DFT calculations. Large dark-gray spheres - Ag(II), small gray spheres - F, green spheres - Li. (top) Ag -- F sublattice, (middle) Li -- F sublattice, and (bottom) polyhedral coordination spheres are shown.}
    \label{F3} 
\end{figure}

Within the Li$_2$AgF$_4$ stoichiometry (Figure \ref{F3}), the Li$_2$AgF$_4\_1$ polymorph is predicted to be the most stable. It crystallises in the common centrosymmetric $C$2/c space group. Again, Ag(II) adopts a severely elongated octahedral environment typical of strong Jahn-Teller distortion. The four shorter Ag–-F bonds range from 2.066\AA\ to 2.071\AA, while the two longer ones are about 2.73\AA. The angle of Ag-F-Ag between the short bonds is 130.0$^\circ$. Lithium cations are coordinated by nearly regular octahedra, with six Li–F distances in the 1.946\AA–2.094\AA\ range.

The structure of Li$_2$AgF$_4\_3$ is approximately 9.27\% larger in volume than Li$_2$AgF$_4\_2$. The first of those  belongs to the most common monoclinic $P2_1$/c space group, whereas the second one to a rare lower-symmetry $P2_1$ one. The latter may give rise to diverse useful nonlinear properties typical of non-centrosymmetric crystals. In both cases, the Ag(II) cations adopt an elongated octahedral coordination environment. In  Li$_2$AgF$_4\_2$, two types of short Ag–F bonds are observed (2.058\AA\ and 2.072\AA) together with two very long ones (2.934\AA). For  Li$_2$AgF$_4\_3$, four short Ag–F bonds range from 2.059\AA\ to 2.086\AA, which are slightly longer than those in Li$_2$AgF$_4\_2$, whereas the two longer Ag–F bonds (2.484\AA\ and 2.526\AA) are somewhat shorter. The fact they differ from each other certifies the lack of the symmetry center at Ag(II) site; this has also been observed for the \textit{HP1} structure of AgF$_2$ at elevated pressure \cite{54}. In  Li$_2$AgF$_4\_2$ the short Ag-F bonds are almost perpendicular (98.7$^{\circ}$ angle) and in  Li$_2$AgF$_4\_3$ the Ag-F-Ag angle is 134.0$^{\circ}$.

The lithium coordination environment in  Li$_2$AgF$_4\_2$ can be described as an elongated octahedron with four short Li–F bonds (1.933\AA–1.962\AA) and two longer ones (2.043\AA\ and 2.113\AA). In contrast,  Li$_2$AgF$_4\_3$ exhibits a slightly distorted tetrahedral coordination, with Li–F bond lengths ranging from 1.878\AA\ to 1.966\AA. Such bond length difference for octahedral and tetrahedral coordination spheres is in line with the values of ionic radii of Li$^{+}$ in these environments (0.73\AA\ for CN=4, 0.90\AA\ for CN=6).

Previously studied M$_2$AgF$_4$ compounds are characterized either by a layered perovskite structure with [AgF$_{4/2+2}$]$^{2-}$ layers, or by a post-perovskite (Na$_2$CuF$_4$-type) structures featuring chains of elongated [AgF$_6$]$^{4-}$ octahedra \cite{38}. In contrast (but similarly to the LiAgF$_3$ phases), for the Li$_2$AgF$_4$ structures explored in this work, tilted $[AgF_4]^{2-}$ chains are present in most cases, except for  Li$_2$AgF$_4\_2$, where isoleted squares [AgF$_4$]$^{2-}$ are found (similar to those in BaAgF$_4$ \cite{55}). The preference for isolated squares over kinked chains may be explained by the higher Lewis acidity of Li$^+$ compared to Na$^+$, which promotes stronger interactions with the fluoride ligands.

\subsection{Magnetic properties}

For all cells studied the magnetic cells are equivalent to the unit cells. Structural relaxations were carried out under both ferromagnetic (FM) and antiferromagnetic (AFM) spin orderings, with AFM proving more stable in all cases (methodology and corresponding illustrations are available in the \textit{Supplementary Information}). Spin-polarised DFT+U calculations further indicate that magnetic ordering has only a minor effect on the relative energies of the polymorphs. 

We have derived the magnetic superexchange constants, J, for all structures. For LiAgF$_3\_1$, the calculated superexchange constant has a value of $J = -78$~meV. Among the Li$_2$AgF$_4$ stoichiometries, the highest $J$ value was calculated for Li$_2$AgF$_4\_3$, with $J = -95$~meV, which is comparable with that found experimentally for KAgF$_3$ \cite{61}. On the other hand, for Li$_2$AgF$_4\_2$ the superexchange constant was the smallest among all investigated structures: $J = -4$~meV. This arises from the silver(II) cations being arranged almost perpendicularly to each other and the Ag--F--Ag superexchange pathway forming between one very long and one short Ag--F bond, which strongly suppresses the magnetic coupling. The Li$_2$AgF$_4\_1$ superexchange constant had a value of $J = -62$~meV due to the presence of the Ag-F-Ag chain in this structure.

\begin{figure}[H]
    \centering
\includegraphics[width=1\textwidth]{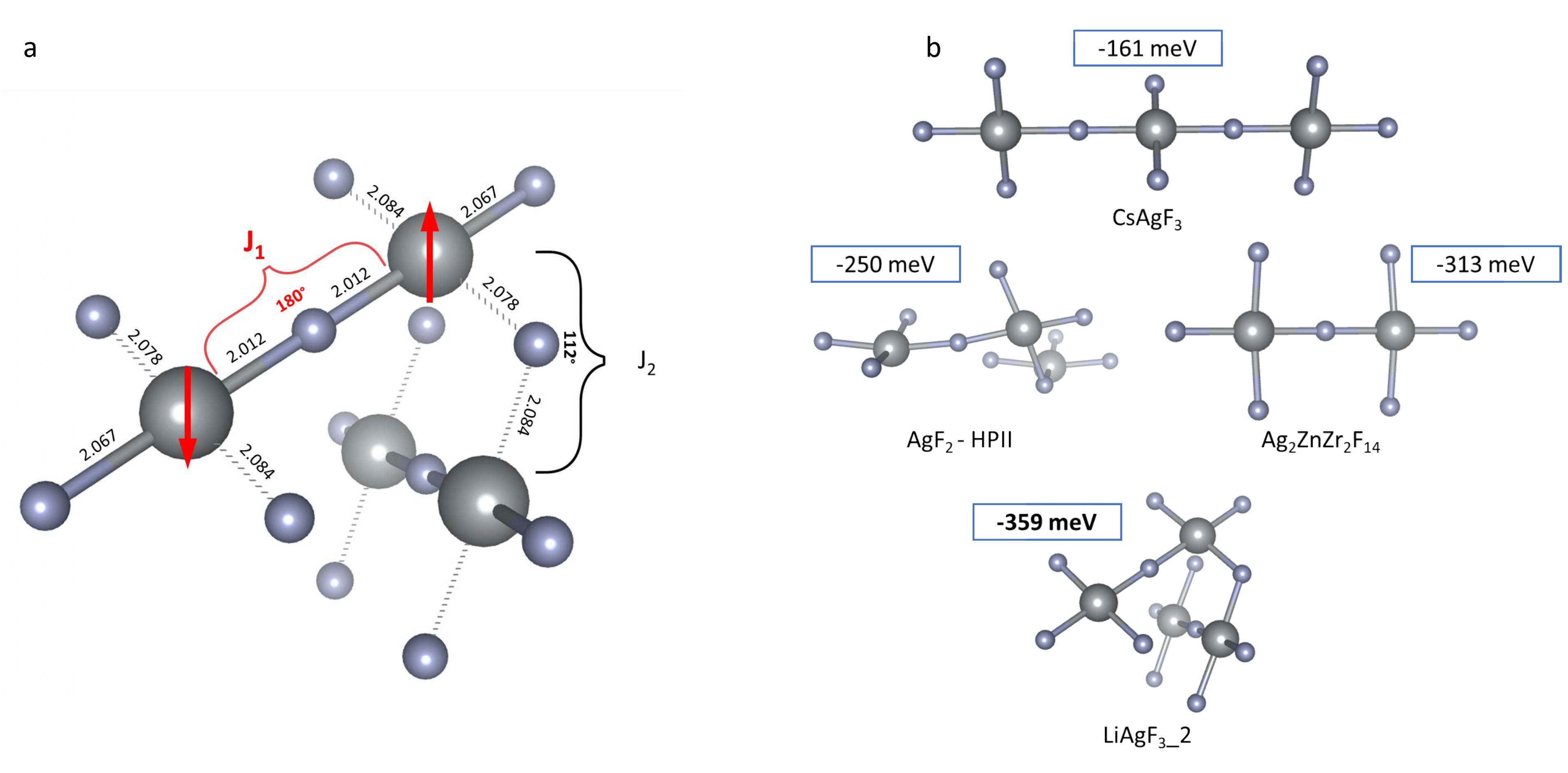} 
\caption{a - Local magnetic structure of LiAgF$_3\_1$ showing the orientations of the spin moments on the Ag(II) cations. The solid lines indicate the nearest-neighbor superexchange pathways ($J_1$), and the dashed blue lines the next-nearest-neighbor ones ($J_2$); bond lengths and angles are shown; b - Comparison of different structural motifs containing the \(\mathrm{[Ag_2F_7]}\) unit, including chain- and dimer-type arrangements, together with their corresponding magnetic superexchange constants.}
    \label{F5} 
\end{figure}

LiAgF$_3\_2$ is a remarkable exception here as it exhibits an immense superexchange constant of $J_1 = -359$ meV within the $[Ag_2F_7]$ dimers and a much smaller one of $J_2 = -11$ meV between them (Figure \ref{F5}a). The [AgF$_4$] squares are interconnected via very short Ag--F bonds (2.014\AA) at a 180$^\circ$ angle, which enables this exceptionally strong superexchange. Other structures exhibiting similar structural motifs have been reported, such as CsAgF$_3$, AgF$_2$-\textit{HPII} and Ag$_2$ZnZr$_2$F$_{14}$ (Figure \ref{F5}b) which also exhibit high $J$ values ($-161$ meV, $-250$ meV and $-313$ meV, respectively, calculated using identical methodology as the one used here) \cite{56}. The value calculated for LiAgF$_3\_2$ is  larger than that previously measured for Sr$_2$CuO$_3$ (-- 240 meV) \cite{60}.

This leads to the conclusion that the LiAgF$_3\_2$ polymorph is worth targetting in experiment, as it may host a record-breaking superexchange between transition metal cations.

\subsection{Thermodynamic stability analysis and synthesis pathways}

\begin{figure}[H]
    \centering
    \includegraphics[width=1\textwidth]{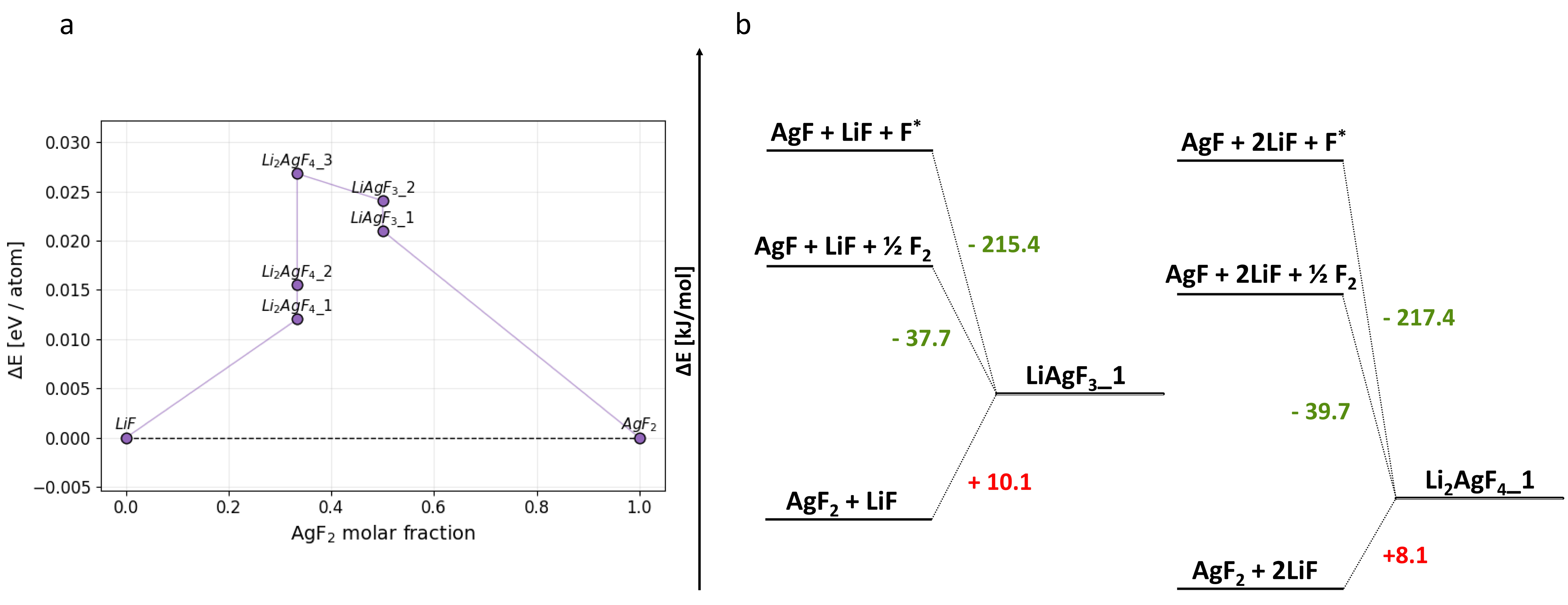} 
    \caption{a - Convex hull of the investigated systems as a function of the AgF$_2$ molar fraction. All structures lie slightly above the convex hull, indicating that they are thermodynamically unstable with respect to binary fluorides; b - Schematic energy profiles ($\Delta E$ [kJ/mol]) for the formation of the most stable lithium–silver(II)–fluorine compounds. The left diagram corresponds to LiAgF$_3\_1$, and the right one to Li$_2$AgF$_4\_1$. The figure indicates that the synthesis of the presented structures is feasible, though it may proceed via alternative substrates (i.e., not  from AgF$_2$ and LiF).}
    \label{F4} 
\end{figure}

The calculations revealed that even the most energetically favourable Li–-Ag(II)–-F arrangements are slightly unstable (by 8-18 kJ/mol) with respect to decomposition into known binary phases.  All candidate polymorphs considered here lie above the convex hull (Figure~\ref{F4}a), indicating their metastability with respect to LiF and AgF$_2$ mixture. To assess whether temperature entropic effects could stabilise any of these phases, the entropy contribution was estimated following the approach proposed by Jenkins and Glasser~\cite{35}. The corresponding stabilisation temperatures, at which the free-energy penalty ($\Delta E_{\text{form}} - T\Delta S$) would be null, are summarised in Table~\ref{T1}.

The calculated stabilization temperatures are quite high and range from $\approx 900$~K for $Li_2AgF_4\_3$ to $\approx 2400$~K  for Li$_2$AgF$_4\_1$. Noting that AgF$_2$ begins to decompose thermally at approximately 600~K (ca.\ 300~$^\circ$C) under dynamic vacuum or inert atmosphere conditions~\cite{40,41}, these results imply that none of the proposed structures could be obtained via elevated temperature synthesis.

\begin{table}[h]
    \centering
    \caption{Thermodynamic properties of the investigated systems, including the internal energy difference $\Delta E$, volume difference per formula unit $\Delta V$, and estimated absolute entropy difference as compared to binary phases  $\Delta S$. The entropy term was calculated based on the volume differences between the hypothetical mixture of $xAgF_2 + yLiF$ and the DFT+U-optimized structures \cite{35}. The temperature $T$ at which the entropy contribution compensates the internal energy difference is also reported.}
    \label{T1}
    \begin{tabular}{ccccc}
        \toprule
        & $\Delta E$ [kJ/mol] & $\Delta V$ [\AA$^3$/f.u.] & $\Delta S$ [J/mol$\cdot$K] & T [K] \\
        \midrule
        $LiAgF_3\_1$     & 10.1 & 5.7   & 10.1 & 1004  \\
        $LiAgF_3\_2$     & 11.6 & 3.5   & 6.2  & 1866  \\
        $Li_2AgF_4\_1$   & 8.1  & 1.9   & 3.3  & 2449  \\
        $Li_2AgF_4\_2$   & 10.5 & 3.9   & 6.9  & 1512  \\
        $Li_2AgF_4\_3$   & 18.1 & 11.0  & 19.4 & 934   \\
        \bottomrule
    \end{tabular}
\end{table}

Considering that, phases with stabilisation temperatures below the decomposition point of AgF$_2$ could, in principle, be obtained without prior breakdown of the binary fluoride precursor. In this context,  Li$_2$AgF$_4\_3$ (with $T \approx 934$~K) lies very close to the AgF$_2$ decomposition threshold, suggesting marginal synthetic accessibility from the mixture of the binary fluorides. All other predicted phases require significantly higher temperatures, where at such conditions, AgF$_2$ is expected to decompose, releasing F$_2$, and therefore the formation of these ternary compounds becomes difficult under equilibrium synthesis routes.
It is worth noting that although the  Li$_2$AgF$_4\_1$ structure lies closest to the stability line, the corresponding stabilisation temperature presented in Table~\ref{T1} is the highest among the investigated phases. The high stabilization temperature is a result of the smallest difference between the predicted size of the unit cell and the sum of volumes of the substrates, which leads to a reduced entropy contribution.

On the other hand, there should be other possible ways of synthesis to obtain those compounds, such as high-temperature fluorination under fluorine pressure, followed by rapid quenching which could lead to trapping metastable arrangements that are inaccessible through conventional solid state-solid state reactions. To explore whether alternative synthetic routes are possible, we computed the formation energies of reaction pathways involving AgF, LiF, and either F$_2$ (fluorine gas) or F$^*$ (fluorine radical). Both pathways are predicted to be thermodynamically feasible, indicating that the formation of these phases is energetically accessible. The calculated formation energies ($\Delta E$) for all considered LiAgF$_3$ and Li$_2$AgF$_4$ polymorphs are summarised in Figure~\ref{F4}b. As expected, reactions involving the fluorine radical are significantly more exothermic than those with molecular fluorine, suggesting that radical fluorination could facilitate compound formation under suitable conditions. Yet another potential pathway is provided by a gentle thermal decomposition of LiAg$^{(III)}$F$_4$, by analogy to the route employed for the preparation of KAgF$_3$ \cite{61}.

\section{Conclusions}

In this work, we investigated the stability and magnetic properties of hypothetical ternary compounds in the Li--Ag(II)--F system. Convex hull analysis shows that all candidate phases are metastable with respect to decomposition into LiF and AgF$_2$. Although entropic effects could theoretically stabilize these polymorphs at finite temperatures, their experimental realization may be limited by the low thermal instability of AgF$_2$. Nevertheless, our analysis suggests that these Li--Ag--F structures could potentially be accessed via alternative synthetic routes, for instance by combining AgF, LiF, and fluorine gas or fluorine radicals. These findings provide motivation for experimental efforts to explore such metastable compounds, as they may exhibit interesting magnetic and structural properties despite their metastability. Indeed, our results indicate that LiAgF$_3\_2$ polymorph exhibits a record-large superexchange constant of $J = -359$~meV within [Ag$_2$F$_7$] dimers, arising from its very short bridging Ag--F bonds. Although this J is exhibited here by quasi-0D bimetallic dimers, yet it suggests that exceptionally J values could be found also for 1D and 2D Ag(II)-based systems. Thus, LiAgF$_3\_2$ represents a meaningful step towards the regime in which magnetic-fluctuations-driven room-$T_c$ superconductivity is theoretically predicted to emerge, as such behaviour is expected for 2D systems with superexchange strengths on the order of $J \approx 400$--$700$~meV~\cite{58}. 
Overall, these findings establish Li--Ag--F systems as a promising and previously unexplored family of materials, highlight the possibility of achieving unprecedented strong magnetic interactions in Ag(II)-based fluorides, and provide a foundation for future synthetic and theoretical investigations into highly-correlated quantum materials.

\section{Acknowledgements}

This work was supported by the Polish National Science Center (project  2024/53/B/ST5/00631). Computations were
performed at the ICM (University of Warsaw, GA83-34).

\section{Supplementary Information}

The Electronic Supplementary Information file contains crystallographic information files for the lowest energy structures, the magnetic models and derivation of the J values, as well as electronic band structure and density of states together with the values of the direct and indirect band gaps.

\printbibliography

\newpage



\section*{Supplementary material}

# Prediction of Novel Li–Ag(II)–F Compounds using Evolutionary Algorithms - Supplementary information

Katarzyna Kuder, Wojciech Grochala

# Contents

# S1    Plane-Wave Cutoff Energy ($E_{cut}$) and k-Point Mesh Convergence

The convergence of the total energy was examined as a function of the plane-wave cutoff energy. As shown in Figure S1, the energy difference between successive cutoff values becomes negligible above approximately 500 eV. Therefore, a cutoff energy of 600 eV was employed for all calculations as it provides a balance between accuracy and computational cost.

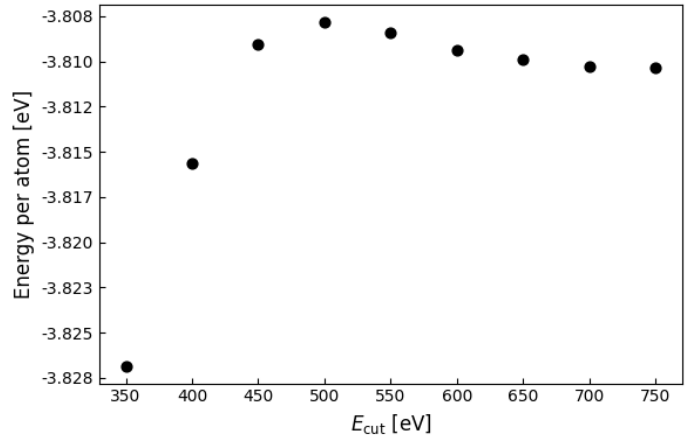

Figure S1: The convergence of the total energy as a function of the plane-wave cutoff energy.

Similarly, the convergence of the total energy as a function of increased k-point sampling was assessed. As presented in Figure S2, the energy variation between successive k-point spacing increments is negligible, therefore, an intermediate k-point spacing of 0.065 Å$^{-1}$ was adopted for all production calculations.

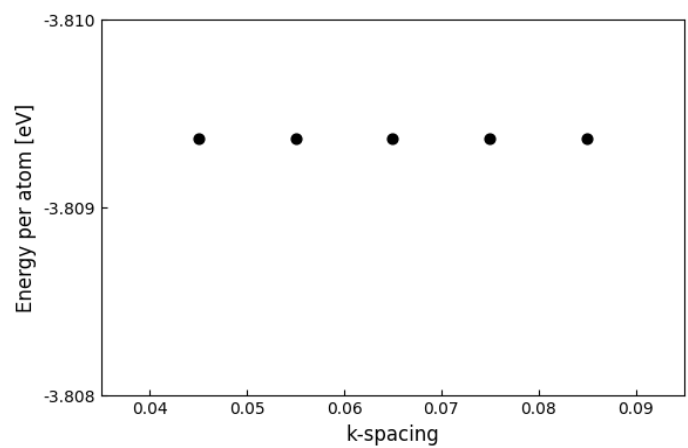

Figure S2: Total energy convergence with respect to k-point sampling.

# S2 Crystallographic Information Files

## S2.1 LiAgF$_3$_1

```
# CIF file created by FINDSYM, version 7.1.3

data_findsym-output
_audit_creation_method FINDSYM

_cell_length_a    3.4101600000
_cell_length_b    14.6280813694
_cell_length_c    4.9988100000
_cell_angle_alpha 90.0000000000
_cell_angle_beta  93.7721300000
_cell_angle_gamma 90.0000000000
_cell_volume      248.8209086510

_symmetry_space_group_name_H-M "C 1 c 1"
_symmetry_Int_Tables_number 9
_space_group.reference_setting '009:C -2yc'
_space_group.transform_Pp_abc a,b,c;0,0,0

loop_
_space_group_symop_id
_space_group_symop_operation_xyz
1 x,y,z
2 x,-y,z+1/2
3 x+1/2,y+1/2,z
4 x+1/2,-y+1/2,z+1/2

loop_
_atom_type_symbol
Ag
F
Li

loop_
_atom_site_label
```

```
_atom_site_type_symbol
_atom_site_symmetry_multiplicity
_atom_site_Wyckoff_symbol
_atom_site_fract_x
_atom_site_fract_y
_atom_site_fract_z
_atom_site_occupancy
_atom_site_fract_symmform
Ag1 Ag   4 a   0.7469825000   0.1642350000   0.2922900000   1.0000000000 Dx,Dy,Dz
F1  F    4 a   0.3179150000   0.0735675000   0.1346650000   1.0000000000 Dx,Dy,Dz
F2  F    4 a   0.3073650000   0.4172975000   0.1273100000   1.0000000000 Dx,Dy,Dz
F3  F    4 a   0.1762750000   0.2519125000   0.4595050000   1.0000000000 Dx,Dy,Dz
Li1 Li   4 a   0.8195775000   0.4490700000   0.2636200000   1.0000000000 Dx,Dy,Dz

# end of cif
```

## S2.2 LiAgF$_3$_2

```
# CIF file created by FINDSYM, version 7.1.3

data_findsym-output
_audit_creation_method FINDSYM

_cell_length_a       5.0789500000
_cell_length_b       5.8001700000
_cell_length_c       8.3159200000
_cell_angle_alpha   94.8966500000
_cell_angle_beta   100.4164200000
_cell_angle_gamma   90.1959500000
_cell_volume       240.0157193444

_symmetry_space_group_name_H-M "P -1"
_symmetry_Int_Tables_number 2
_space_group.reference_setting '002:-P 1'
_space_group.transform_Pp_abc a,b,c;0,0,0

loop_
_space_group_symop_id
_space_group_symop_operation_xyz
1 x,y,z
2 -x,-y,-z

loop_
_atom_type_symbol
Ag
F
Li

loop_
_atom_site_label
_atom_site_type_symbol
_atom_site_symmetry_multiplicity
_atom_site_Wyckoff_symbol
_atom_site_fract_x
_atom_site_fract_y
_atom_site_fract_z
_atom_site_occupancy
```

```
_atom_site_fract_symmform
Ag1  Ag   2 i   0.3114900000   0.1138500000   0.6836900000   1.0000000000 Dx,Dy,Dz
Ag2  Ag   2 i   0.8125300000   0.6124700000   0.6742600000   1.0000000000 Dx,Dy,Dz
F1   F    1 f   0.5000000000   0.0000000000   0.5000000000   1.0000000000 0,0,0
F2   F    2 i   0.3920800000   0.0139400000   0.1428400000   1.0000000000 Dx,Dy,Dz
F3   F    2 i   0.3633300000   0.5173100000   0.1434300000   1.0000000000 Dx,Dy,Dz
F4   F    2 i   0.8581800000   0.7647500000   0.1206900000   1.0000000000 Dx,Dy,Dz
F5   F    2 i   0.0221650000   0.2527300000   0.5097000000   1.0000000000 Dx,Dy,Dz
F6   F    1 h   0.5000000000   0.5000000000   0.5000000000   1.0000000000 0,0,0
F7   F    2 i   0.8842500000   0.2635300000   0.1374350000   1.0000000000 Dx,Dy,Dz
Li1  Li   2 i   0.5135900000   0.2567600000   0.0271400000   1.0000000000 Dx,Dy,Dz
Li2  Li   1 c   0.0000000000   0.5000000000   0.0000000000   1.0000000000 0,0,0
Li3  Li   1 a   0.0000000000   0.0000000000   0.0000000000   1.0000000000 0,0,0

# end of cif
```

## S2.3 $Li_2AgF_{4-1}$

```
# CIF file created by FINDSYM, version 7.1.3

data_findsym-output
_audit_creation_method FINDSYM

_cell_length_a      20.9862000000
_cell_length_b       5.7389300000
_cell_length_c       5.0092000000
_cell_angle_alpha   90.0000000000
_cell_angle_beta    99.4446700000
_cell_angle_gamma   90.0000000000
_cell_volume       595.1216834805

_symmetry_space_group_name_H-M "C 1 2/c 1"
_symmetry_Int_Tables_number 15
_space_group.reference_setting '015:-C 2yc'
_space_group.transform_Pp_abc a,b,c;0,0,0

loop_
_space_group_symop_id
_space_group_symop_operation_xyz
1 x,y,z
2 -x,y,-z+1/2
3 -x,-y,-z
4 x,-y,z+1/2
5 x+1/2,y+1/2,z
6 -x+1/2,y+1/2,-z+1/2
7 -x+1/2,-y+1/2,-z
8 x+1/2,-y+1/2,z+1/2

loop_
_atom_type_symbol
Ag
F
Li

loop_
_atom_site_label
```

```
_atom_site_type_symbol
_atom_site_symmetry_multiplicity
_atom_site_Wyckoff_symbol
_atom_site_fract_x
_atom_site_fract_y
_atom_site_fract_z
_atom_site_occupancy
_atom_site_fract_symmform
Ag1  Ag   8 f   0.4326900000   0.2508600000  -0.0462000000   1.0000000000 Dx,Dy,Dz
F1   F    8 f   0.3589900000   0.1207075000   0.6683000000   1.0000000000 Dx,Dy,Dz
F2   F    8 f   0.2498600000   0.3738700000   0.7476600000   1.0000000000 Dx,Dy,Dz
F3   F    4 e   0.0000000000  -0.0972100000   0.2500000000   1.0000000000 0,Dy,0
F4   F    4 e   0.0000000000   0.4013200000   0.2500000000   1.0000000000 0,Dy,0
F5   F    8 f   0.3595500000   0.3790200000   0.1443900000   1.0000000000 Dx,Dy,Dz
Li1  Li   8 f   0.8020800000   0.3794500000   0.4553900000   1.0000000000 Dx,Dy,Dz
Li2  Li   8 f   0.3100150000   0.3676500000   0.4530300000   1.0000000000 Dx,Dy,Dz

# end of cif
```

## S2.4 $Li_2AgF_4\_2$

```
# CIF file created by FINDSYM, version 7.1.3

data_findsym-output
_audit_creation_method FINDSYM

_cell_length_a     5.4808500000
_cell_length_b     5.7988100000
_cell_length_c     5.0321900000
_cell_angle_alpha  90.0000000000
_cell_angle_beta   107.0774600000
_cell_angle_gamma  90.0000000000
_cell_volume       152.8833540110

_symmetry_space_group_name_H-M "P 1 21/c 1"
_symmetry_Int_Tables_number 14
_space_group.reference_setting '014:-P 2ybc'
_space_group.transform_Pp_abc a,b,c;0,0,0

loop_
_space_group_symop_id
_space_group_symop_operation_xyz
1 x,y,z
2 -x,y+1/2,-z+1/2
3 -x,-y,-z
4 x,-y+1/2,z+1/2

loop_
_atom_type_symbol
Ag
F
Li

loop_
_atom_site_label
_atom_site_type_symbol
```

```
_atom_site_symmetry_multiplicity
_atom_site_Wyckoff_symbol
_atom_site_fract_x
_atom_site_fract_y
_atom_site_fract_z
_atom_site_occupancy
_atom_site_fract_symmform
Ag1 Ag   2 d   0.5000000000   0.0000000000   0.5000000000   1.0000000000 0,0,0
F1  F    4 e   0.7910950000   0.6299700000   0.8628200000   1.0000000000 Dx,Dy,Dz
F2  F    4 e   0.7871675000   0.3722500000   0.3337975000   1.0000000000 Dx,Dy,Dz
Li1 Li   4 e   0.0304925000   0.8823725000   0.7674950000   1.0000000000 Dx,Dy,Dz

# end of cif
```

### S2.5  $Li_2AgF_4$-3

```
# CIF file created by FINDSYM, version 7.1.3

data_findsym-output
_audit_creation_method FINDSYM

_cell_length_a      3.2936000000
_cell_length_b      5.2045700000
_cell_length_c      9.7778550969
_cell_angle_alpha  90.0000000000
_cell_angle_beta   94.6339649251
_cell_angle_gamma  90.0000000000
_cell_volume      167.0618710437

_symmetry_space_group_name_H-M "P 1 21 1"
_symmetry_Int_Tables_number 4
_space_group.reference_setting '004:P 2yb'
_space_group.transform_Pp_abc a,b,c;0,0,0

loop_
_space_group_symop_id
_space_group_symop_operation_xyz
1 x,y,z
2 -x,y+1/2,-z

loop_
_atom_type_symbol
Ag
F
Li

loop_
_atom_site_label
_atom_site_type_symbol
_atom_site_symmetry_multiplicity
_atom_site_Wyckoff_symbol
_atom_site_fract_x
_atom_site_fract_y
_atom_site_fract_z
_atom_site_occupancy
_atom_site_fract_symmform
Ag1 Ag   2 a   0.7815300000   0.8254950000   0.3706100000   1.0000000000 Dx,Dy,Dz
```

```
F1    F     2 a   0.7569450000   0.8231950000  -0.0209550000   1.0000000000 Dx,Dy,Dz
F2    F     2 a   0.6658400000   0.4935350000   0.7630700000   1.0000000000 Dx,Dy,Dz
F3    F     2 a   0.7903850000   0.5034750000   0.2485350000   1.0000000000 Dx,Dy,Dz
F4    F     2 a   0.2199000000   0.6470150000   0.5042900000   1.0000000000 Dx,Dy,Dz
Li1   Li    2 a   0.7295550000   0.4603400000  -0.0360100000   1.0000000000 Dx,Dy,Dz
Li2   Li    2 a   0.2840450000   0.3533400000   0.2122100000   1.0000000000 Dx,Dy,Dz

# end of cif
```

# S3 Magnetic Configurations

Structural relaxations were performed for both ferromagnetic (FM) and antiferromagnetic (AFM) spin configurations. In all cases, the AFM arrangement was found to be more stable. In the figures below, the magnetic models of all structures are presented.

The superexchange constants were calculated from:

$$H = -\frac{1}{2}\sum_{ij} J_{ij} S_i S_j \qquad (1)$$

where $J_{ij}$ is the magnetic coupling constants between spin sites. Positive values of $J_{ij}$ correspond to FM spin ordering and negative $J_{ij}$ values correspond to AFM spin ordering.

For each structure listed in the tables below, we report the model energy expressions, spin ordering along with the superexchange constants extracted from our calculations.

Table S1: Spin configurations and corresponding energy expressions for the LiAgF$_3$_1 models.

| Model | Atom 1 | 2 | 3 | 4 | Energy |
|---|---|---|---|---|---|
| FM | + | + | + | + | $E = E_0 - J$ |
| AFM | + | − | + | − | $E = E_0 + J$ |
| | | J [meV]: | | | −77.74 |

Table S2: Spin configurations and corresponding energy expressions for the LiAgF$_3$_2 models.

| Model | Atom 1 | 2 | 3 | 4 | Energy |
|---|---|---|---|---|---|
| FM | + | + | + | + | $E = E_0 - 0.5J_1 - 0.5J_2$ |
| AFM | + | − | + | − | $E = E_0 + 0.5J_1 + 0.5J_2$ |
| AFM2 | − | + | + | − | $E = E_0 + 0.5J_1 - 0.5J_2$ |
| | | J$_1$ [meV]: | | | −358.75 |
| | | J$_2$ [meV]: | | | −10.54 |

Table S3: Spin configurations and corresponding energy expressions for the Li$_2$AgF$_4$_1 models.

| Model | Atom 1 | 2 | 3 | 4 | 5 | 6 | 7 | 8 | Energy |
|---|---|---|---|---|---|---|---|---|---|
| FM | + | + | + | + | + | + | + | + | $E = E_0 - 2J$ |
| AFM | + | − | − | + | + | − | − | + | $E = E_0 + 2J$ |
| | | | | J [meV]: | | | | | −62.46 |

Table S4: Spin configurations and corresponding energy expressions for the Li$_2$AgF$_4$_2 models.

| Model | Atom 1 | Atom 2 | Energy |
|---|---|---|---|
| FM | + | + | $E = E_0 - 0.5J$ |
| AFM | − | + | $E = E_0 + 0.5J$ |
| J [meV]: | | | −4.07 |

Table S5: Spin configurations and corresponding energy expressions for the Li$_2$AgF$_4$_3 models.

| Model | Atom 1 | Atom 2 | Energy |
|---|---|---|---|
| FM | + | + | $E = E_0 - 0.5J$ |
| AFM | − | + | $E = E_0 + 0.5J$ |
| J [meV]: | | | −95.48 |

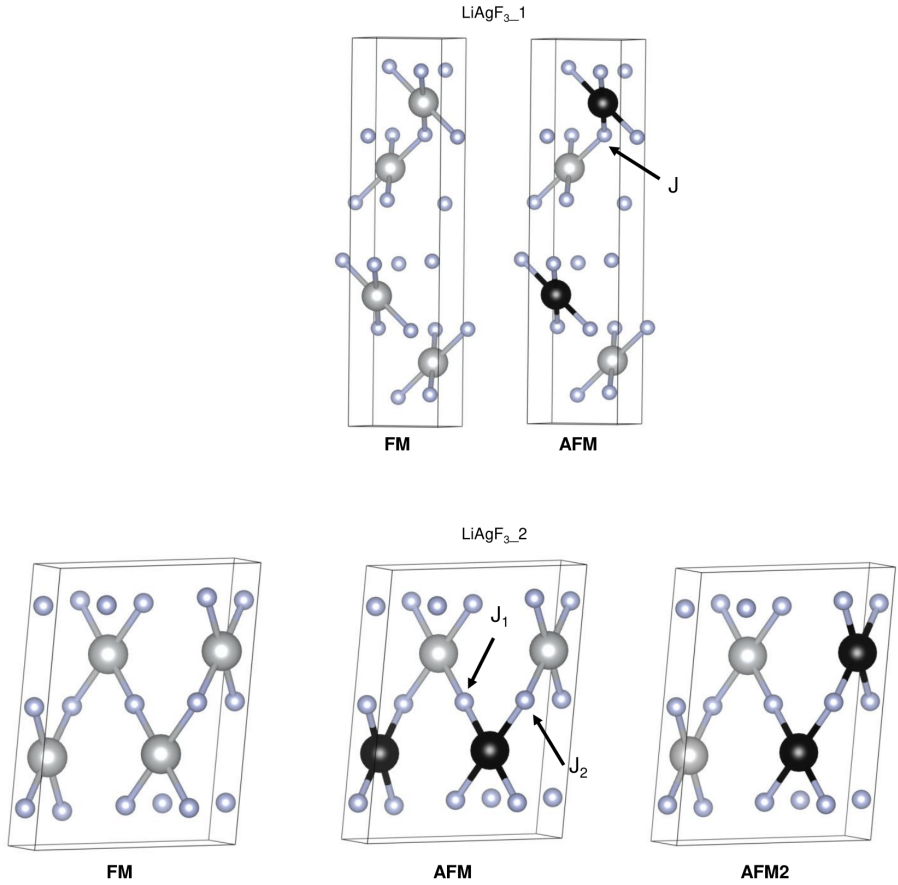

Figure S3: Magnetic configurations of all investigated LiAgF$_3$ systems are shown. Spin-up Ag(II) cations are represented by grey spheres, while spin-down cations are shown as black spheres. For clarity, Li atoms have been omitted from the illustrations. The calculated superexchange constants are indicated on the corresponding ground-state magnetic structures.

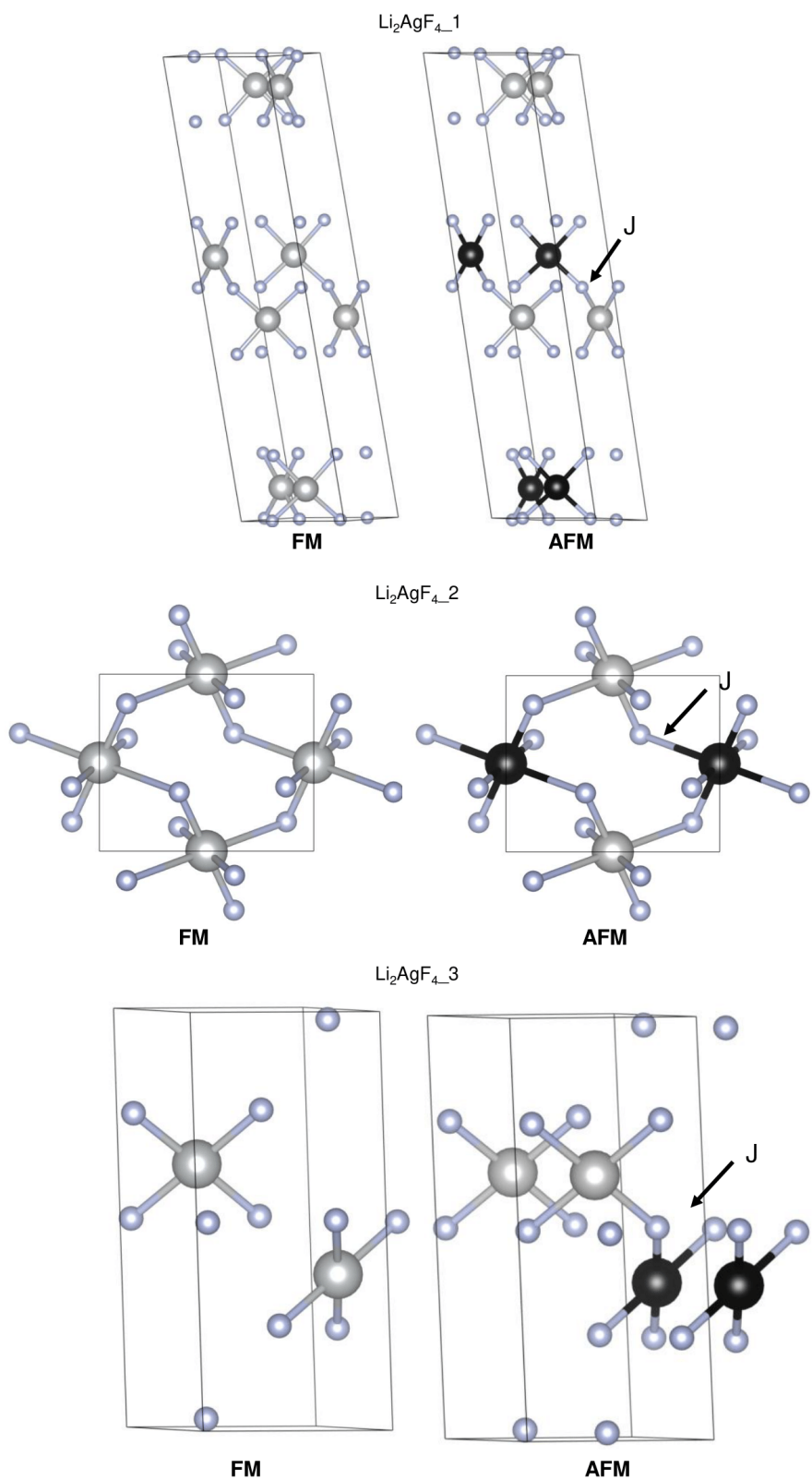

Figure S4: Magnetic configurations of all investigated $Li_2AgF_4$ systems are shown. Spin-up Ag(II) cations are represented by grey spheres, while spin-down cations are shown as black spheres. For clarity, Li atoms have been omitted from the illustrations. The calculated superexchange constants are indicated on the corresponding ground-state magnetic structures.

9

# S4 Electronic Band Structure and Density of States

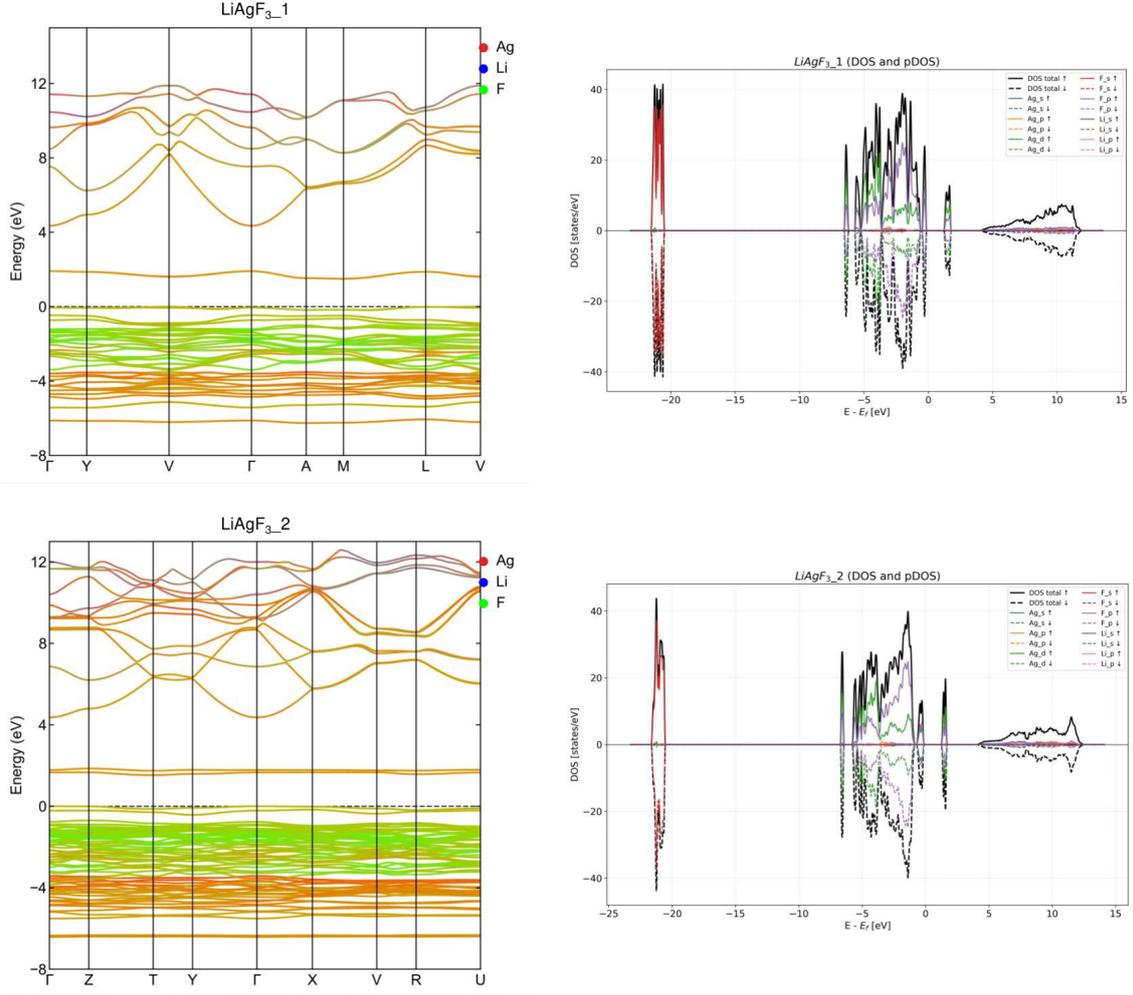

Figure S5: Spin-resolved density of states (DOS) and projected DOS (pDOS) with orbital contributions for the LiAgF$_3$ structures. Spin-up states are shown in the positive direction with solid lines, while spin-down states are shown in the negative direction with dotted lines. The Fermi level is set to 0 eV. For the electronic band structure, an RGB scheme was used, where the line color reflects the character of the band. Each element contributes either red, green, or blue, and the resulting line colour is a mixture of the corresponding contributions.

Table S6: Calculated direct and indirect band gaps for LiAgF$_3$ structures.

| Structure | Direct band gap (eV) | Indirect band gap (eV) |
|---|---|---|
| LiAgF$_3$_1 | 1.608 | 1.499 |
| LiAgF$_3$_2 | 1.656 | 1.523 |

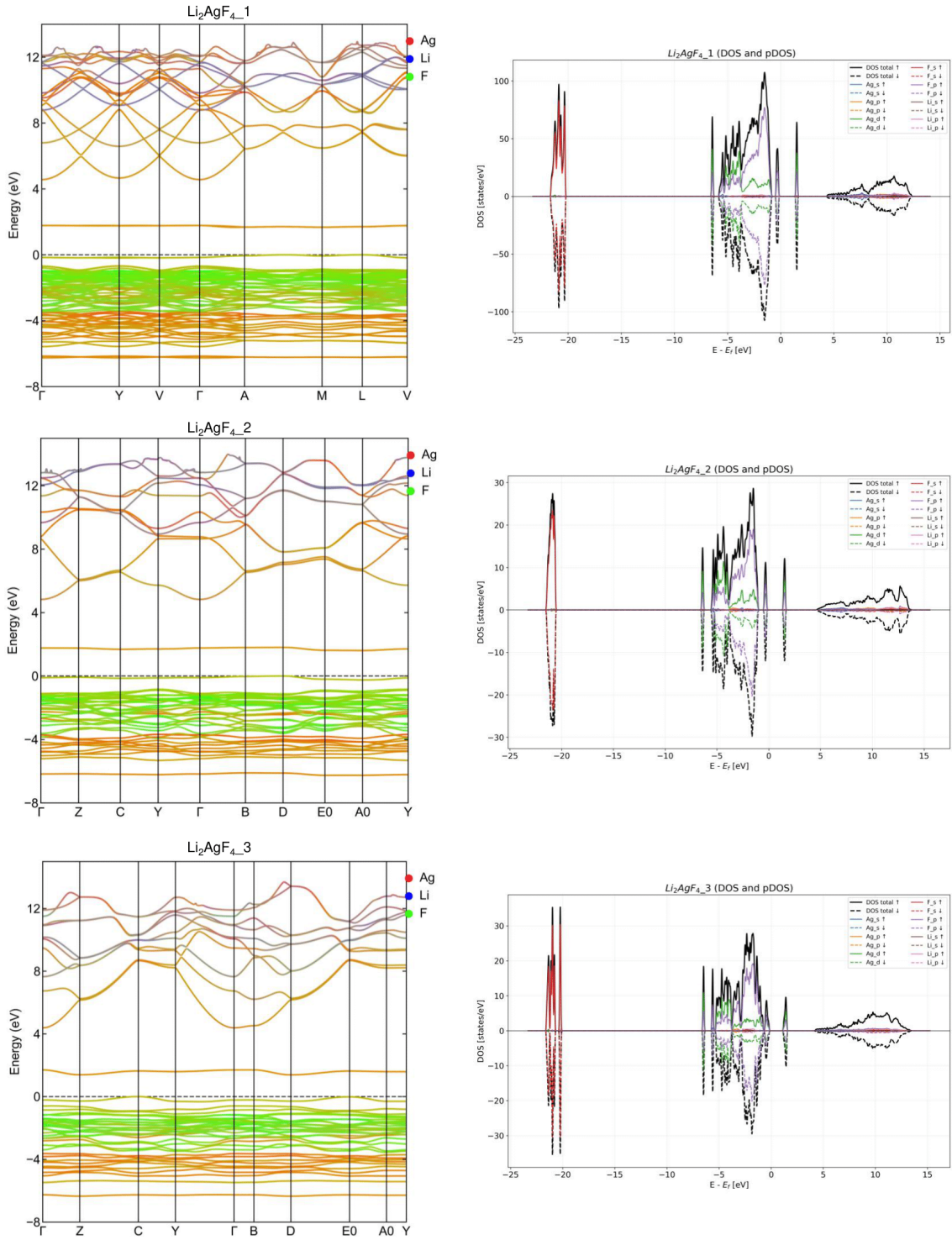

Figure S6: Spin-resolved density of states (DOS) and projected DOS (pDOS) with orbital contributions for the $Li_2AgF_4$ structures. Spin-up states are shown in the positive direction with solid lines, while spin-down states are shown in the negative direction with dotted lines. The Fermi level is set to 0 eV. For the electronic band structure, an RGB scheme was used, where the line color reflects the character of the band. Each element contributes either red, green, or blue, and the resulting line color is a mixture of the corresponding contributions.

Table S7: Calculated direct and indirect band gaps for $Li_2AgF_4$ structures.

| Structure | Direct band gap (eV) | Indirect band gap (eV) |
|---|---|---|
| $Li_2AgF_4$-1 | 1.701 | 1.666 |
| $Li_2AgF_4$-2 | 1.807 | 1.609 |
| $Li_2AgF_4$-3 | 1.631 | 1.385 |



\end{document}